\documentclass[runningheads]{llncs}
\usepackage[T1]{fontenc}
\usepackage{xcolor}
\usepackage{graphicx}
\usepackage{amssymb}
\usepackage{amsmath}
\usepackage{bm}
\usepackage{pifont}
\usepackage{multirow}
\usepackage{booktabs}
\usepackage{subfigure}
\usepackage{makecell}
\usepackage{xurl}
\usepackage{xcolor}
\usepackage[export]{adjustbox}
\usepackage{graphicx}
\usepackage[colorlinks,
            linkcolor=blue,
            anchorcolor=blue,
            citecolor=blue]{hyperref}
\usepackage{graphicx,verbatim}
\usepackage{marvosym}
\begin{document}
\title{Taming Stable Diffusion for Computed Tomography Blind Super-Resolution}
%

\author{Chunlei Li\inst{1}\thanks{Equal contribution.} \and
Yilei Shi\inst{1}$^\star$ \and
Haoxi Hu\inst{2} \and
Jingliang Hu\inst{1} \and
Xiao Xiang Zhu\inst{3} \and
Lichao Mou\inst{1}\textsuperscript{(\Letter)}}

\authorrunning{C. Li et al.}
%
\institute{MedAI Technology (Wuxi) Co. Ltd., Wuxi, China\\\email{lichao.mou@medimagingai.com} \and Northeastern University, Boston, USA \and Technical University of Munich, Munich, Germany}

\maketitle              
\begin{abstract}
High-resolution computed tomography (CT) imaging is essential for medical diagnosis but requires increased radiation exposure, creating a critical trade-off between image quality and patient safety. While deep learning methods have shown promise in CT super-resolution, they face challenges with complex degradations and limited medical training data. Meanwhile, large-scale pre-trained diffusion models, particularly Stable Diffusion, have demonstrated remarkable capabilities in synthesizing fine details across various vision tasks. Motivated by this, we propose a novel framework that adapts Stable Diffusion for CT blind super-resolution. We employ a practical degradation model to synthesize realistic low-quality images and leverage a pre-trained vision-language model to generate corresponding descriptions. Subsequently, we perform super-resolution using Stable Diffusion with a specialized controlling strategy, conditioned on both low-resolution inputs and the generated text descriptions. Extensive experiments show that our method outperforms existing approaches, demonstrating its potential for achieving high-quality CT imaging at reduced radiation doses. Our code will be made publicly available.


\end{abstract}
\section{Introduction}
\label{sec:intro}
Computed tomography (CT) is an indispensable imaging modality in modern healthcare, serving a crucial role in disease diagnosis, treatment planning, and clinical research~\cite{miccai_01}. The diagnostic value of CT images heavily depends on their spatial resolution, driving the continuous pursuit of high-resolution CT imaging techniques~\cite{bg_02}. However, achieving high-resolution CT scans often requires increased radiation exposure, which poses potential health risks to patients~\cite{bg_03}. This fundamental trade-off between image quality and radiation dose has motivated extensive research in CT super-resolution methods, aiming to generate high-resolution CT images from low-dose acquisitions~\cite{bg_04}.
\par
Recent years have witnessed remarkable progress in image super-resolution, primarily driven by deep learning approaches. Early efforts~\cite{srgan,rcan,edsr} typically rely on simulated training data with simplified, uniform degradation models, such as bicubic downsampling. Although these methods show promise in controlled settings, they often underperform when applied to real-world low-resolution images characterized by complex, unknown degradations. To address these limitations, blind image super-resolution methods~\cite{real-esrgan,ikc,dan,realsr} have gained increasing attention, particularly in medical imaging applications where degradation patterns can be highly variable and unpredictable. Recent work has demonstrated success in radiographs and MRI through blind super-resolution approaches that generalize across degradation types~\cite{blind_medical1,blind_medical2_mri}.
\par
Despite these remarkable advances, the performance of learning-based super-resolution methods is inherently tied to the availability of high-quality training data. Unlike natural image domains with abundant data, medical imaging faces data scarcity due to privacy concerns and regulatory requirements, creating a barrier to developing robust super-resolution models.
\par
Recently, large-scale pre-trained text-to-image diffusion models, particularly Stable Diffusion~\cite{stablediffusion}, have demonstrated astonishing capabilities in generating highly detailed images. Trained on vast collections of image-text pairs, these models have not only revolutionized image generation but have also shown promise results when adapted to various vision tasks, including image segmentation and enhancement~\cite{sd_seg1,sd_seg2,sd_enhen1,sd_enhen2}. This success motivates our investigation into leveraging Stable Diffusion for CT super-resolution, utilizing its rich prior knowledge of image structure and detail synthesis.
\par
However, applying Stable Diffusion to CT super-resolution presents challenges. A fundamental obstacle is the domain gap between the model's training data---comprising internet-scale natural image-text pairs---and CT images. Direct application of Stable Diffusion risks synthesizing features characteristic of natural images, which is problematic given CT images' distinct properties in grayscale distributions, structural patterns, and texture characteristics.
\par
To address this issue, we propose a novel framework that adapts Stable Diffusion for CT blind super-resolution. First, we employ a practical degradation model~\cite{real-esrgan} to synthesize realistic low-quality images. Second, we utilize LLaVA-Med~\cite{llava-med}, a specialized medical vision-language model, to generate anatomical descriptions from these degraded images, which serve as semantic prompts for the subsequent super-resolution process. Third, we adapt Stable Diffusion by freezing its pre-trained layers and introducing a side-controlling strategy to generate high-resolution CT images conditioned on low-resolution inputs and text prompts. Our approach effectively bridges the gap between Stable Diffusion's general-purpose generative capabilities and specialized medical imaging requirements. Our main contributions are:
\begin{itemize}
\item We present the first attempt to adapt Stable Diffusion for medical image super-resolution.
\item We propose a two-stage framework combining LLaVA-Med with Stable Diffusion, enabling text-guided super-resolution for medical imaging, along with a side-controlling approach to address domain shift.
\item State-of-the-art performance on CT super-resolution tasks, surpassing existing methods.
\end{itemize}

\begin{figure}[t]
\includegraphics[width=\textwidth]{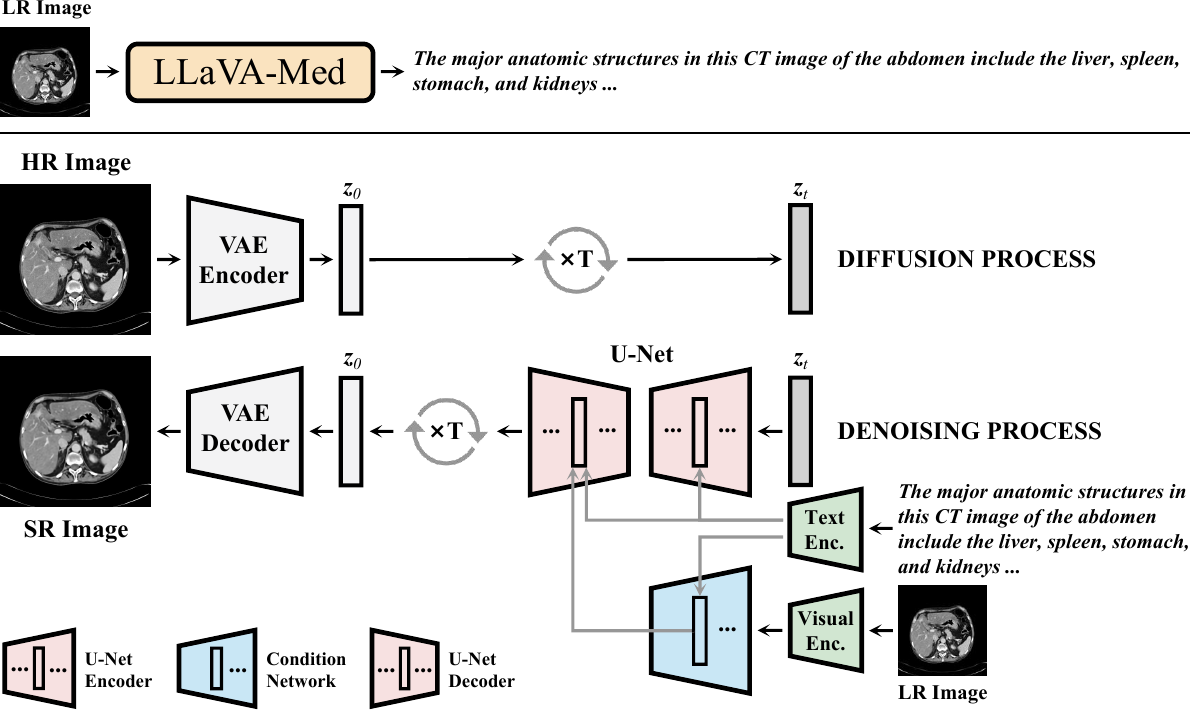}
\caption{Overview of our proposed framework. The pipeline consists of two stages: (1) text description generation from low-resolution images using LLaVA-Med, and (2) image super-resolution through the proposed side-controlling strategy, which integrates both the generated text prompts and low-resolution images into the Stable Diffusion model.}
\label{fig:liuchengtu}
\end{figure}

\section{Method}
\label{sec:method}
As shown in Fig.~\ref{fig:liuchengtu}, our framework consists of two main stages: text prompt generation and taming Stable Diffusion for super-resolution through the side-controlling strategy. The key innovation lies in fusing visual constraints from low-resolution inputs with semantic guidance derived from text prompts, enabling precise control over the pre-trained Stable Diffusion model.

\subsection{Text Prompt Generation}
\label{sec:prompt}
To make full use of the generative potential of Stable Diffusion for super-resolution, we employ LLaVA-Med~\cite{llava-med} to generate descriptive text prompts from low-dose CT images.
\par
Given a low-resolution image $\bm{x}_{lr}$ and an instruction $\bm{s}$ (e.g., \textit{Describe main anatomical structures present in the image}) as inputs, LLaVA-Med produces a corresponding textual description $\bm{p}$, formulated as:
\begin{equation}
\bm{p}=\mathcal{F}_\text{LLaVA-Med}(\bm{x}_{lr}, \bm{s}) \,.
\end{equation}
\par
The resulting $\bm{p}$ provides valuable semantic information, particularly anatomical structures, which is crucial for medical imaging tasks. This description is then fed into Stable Diffusion as a text prompt, along with $\bm{x}_{lr}$.

\subsection{Side-Controlling}
To bridge the domain gap between natural and medical images while leveraging the strength of Stable Diffusion, we propose a side-controlling strategy. This approach synergizes text prompts (see Section~\ref{sec:prompt}) and low-resolution images to provide complementary guidance during generation. The textual prompts offer high-level contextual understanding, while the low-resolution images introduce spatial and structural constraints. These constraints are injected into the Stable Diffusion model in a side-controlling manner. By jointly utilizing these two types of control signals, our approach ensures that the generation of high-resolution images considers both semantic context and fine-grained textures required in medical imaging.
\par
Our strategy comprises three key components: a text encoder for prompt encoding, a visual encoder that encodes low-resolution images, and a condition network that incorporates the encoded features into the diffusion model to guide the generation process.

\subsubsection{Text Encoder}
We leverage the text encoder of CLIP~\cite{clip} to extract linguistic features, as it is pre-trained on massive image-text pairs. We freeze the text encoder to preserve its robust generalization in image-text alignment. Let $\mathcal{E}_\text{T}$ denote the text encoder, the encoding process can be written as:
\begin{equation}
\bm{f}_{p} = \mathcal{E}_\text{T}(\bm{p}) \,.
\end{equation}

\subsubsection{Visual Encoder}
The visual encoder extracts visual features from low-resolution inputs to guide the generative process. We utilize the pre-trained VAE encoder of Stable Diffusion, denoted as $\mathcal{E}_\text{V}$, to project $\bm{x}_{lr}$ into the VAE's latent space:
\begin{equation}
\bm{f}_{lr} = \mathcal{E}_\text{V}(\bm{x}_{lr}) \,.
\end{equation}

\subsubsection{Condition Network}
The latent representation of a high-resolution image is obtained by the pre-trained VAE encoder, denoted $\bm{z}_0$. Both diffusion and denoising processes occur in the latent space. During diffusion, Gaussian noise with variance $\beta_t \in (0, 1)$ at time $t$ is added to $\bm{z}_0$ to produce the noisy latent:
\begin{equation}
\bm{z}_t = \sqrt{\bar{\alpha}_t}\,\bm{z}_0 + \sqrt{1 - \bar{\alpha}_t}\,\epsilon \,,
\end{equation}
where $\epsilon \sim \mathcal{N}(0, \mathbf{I})$, $\alpha_t = 1 - \beta_t$, and $\bar{\alpha}_t = \prod_{s=1}^{t} \alpha_s$. At sufficiently large $t$, the latent $\bm{z}_t$ approximates a standard Gaussian distribution.
\par
We define the entire U-Net in the Stable Diffusion model as $\mathcal{F}_\text{U-Net}$. In the denoising process, we clone a trainable copy of the U-Net encoder $\mathcal{E}_\text{U}$ as our condition network (denoted as $\mathcal{F}_\text{COND}$), which processes condition information and outputs control signals. This copy strategy provides a good weight initialization for the condition network. Then, the concatenation of the condition $\bm{f}_{lr}$ and the noisy latent $\bm{z}_t$ at time $t$ forms $\bm{z}^\prime_t$. This concatenated representation, along with the textual embedding $\bm{f}_{p}$, are fed into $\mathcal{F}_\text{COND}$. At each block of $\mathcal{F}_\text{COND}$, the output is skip-connected to the corresponding block of the U-Net decoder $\mathcal{D}_\text{U}$ in Stable Diffusion, facilitating the effective integration of the conditional signals.
\par
Meanwhile, $\bm{f}_{p}$ and $\bm{z}_t$ are fed into $\mathcal{F}_\text{U-Net}$. We denote the output of the $i$-th block of $\mathcal{D}_\text{U}$ as $\bm{o}_i$. In the $i$-th block where control information is involved, the calculation proceeds as follows:
\begin{equation}
\bm{o}_i = \mathcal{D}^i_{\text{U}}(\bm{o}_{i-1}, \bm{f}_{p}) + \mathcal{F}_\text{COND}^i(\bm{z}^\prime_t, \bm{f}_{p}) \,.
\end{equation}
\par
The final high-resolution image is generated by mapping the denoised latent $\bm{z}^\prime_0$ through the VAE decoder $\mathcal{D}_\text{V}$:
\begin{equation}
\bm{x}^\prime_{hr} = \mathcal{D}_\text{V}(\bm{z}^\prime_0) \,.
\end{equation}

\subsection{Training Objective}
We train a noise prediction model $\epsilon_\theta$ conditioned on low-resolution images and text prompts:
\begin{equation}
\mathcal{L} 
= \mathbb{E}_{\bm{z}_0, \bm{f}_{lr}, t, \bm{f}_{p}, \epsilon \sim \mathcal{N}}
\Bigl[
\bigl\|
\epsilon \;-\;
\epsilon_{\theta}\bigl(\bm{z}_{t}, t, \bm{f}_{lr}, \bm{f}_{p} \bigr)
\bigr\|_{2}^{2}
\Bigr] \,.
\end{equation}
\section{Experiments}
\subsection{Datasets and Evaluation Metrics}
We evaluate our method on two widely used public CT image datasets: Pancreas~\cite{dataset} and 3D-IRCADb~\cite{dataset}. The Pancreas dataset contains 19,328 CT slices with a resolution of 512$\times$512 pixels from 82 patients. For our experiments, we utilize 5,638 slices from 65 patients for training and 1,421 slices from 17 patients for testing. The 3D-IRCADb dataset comprises 2,823 CT slices (512$\times$512 resolution) from 20 patients, with 1,663 slices from 16 patients for training and 411 slices from 4 patients for testing. Following~\cite{dose}, we standardize the data by clipping Hounsfield unit values to $[-135, 215]$ and simulate low-dose noise by setting the blank flux to 0.5$\times$10$^5$. We adopt the degradation pipeline from Real-ESRGAN~\cite{real-esrgan} to generate low-resolution images at 256$\times$256 and 128$\times$128 pixels, corresponding to upscaling factors of $\times$2 and $\times$4, respectively.
\par
For evaluation, we employ two metrics: peak signal-to-noise ratio (PSNR) and structural similarity index measure (SSIM)~\cite{ssim}. They are widely accepted reference-based fidelity measures for assessing image quality.

\subsection{Implementation Details}
All experiments are conducted using PyTorch on NVIDIA RTX 4090Ti GPUs. We train the model for 150K iterations using the Adam optimizer~\cite{adam} with a learning rate of 5$\times$10$^{-5}$ and a batch size of 2 per GPU across 8 GPUs. During inference, we employ spaced DDPM sampling~\cite{sample} with 50 timesteps.

\begin{table}[t]
\caption{Quantitative comparison with state-of-the-art methods on two public datasets at scale factors of $\times$2 and $\times$4. Results in \textbf{bold} and \underline{underlined} indicate the best and second-best performance, respectively.}
\setlength{\tabcolsep}{1mm}
\centering
\resizebox{1.0\linewidth}{!}{
\begin{tabular}{l|cccc|cccc}
\toprule
& \multicolumn{4}{c|}{\textbf{Pancreas}} & \multicolumn{4}{c}{\textbf{3D-IRCADb}} \\ 
\multirow{-3}{*}{}  & \multicolumn{2}{c}{$\times$2} & \multicolumn{2}{c|}{$\times$4} & \multicolumn{2}{c}{$\times$2} & \multicolumn{2}{c}{$\times$4}  \\ 
\multirow{-3}{*}{}  & PSNR & SSIM & PSNR & SSIM & PSNR & SSIM & PSNR & SSIM \\ \midrule
SRGAN  & 26.8732 & 0.7701 & 25.6810 & 0.7539 & 25.5262 & 0.7425 & 24.5693 & 0.7206 \\
EDSR & 28.8972 & 0.8364 & 27.8862 & 0.8105 & 28.2912 & 0.7973 & 27.1478  & 0.7882 \\
ESRGAN & 27.9114 & 0.8073 & 26.7801 & 0.7818 & 26.6362 & 0.7492 & 25.1963 & 0.7366 \\
RCAN & 29.1973 & 0.8442 & 28.0183 & 0.8236 & 28.5914 & 0.7969 & 27.2284 & 0.7891 \\
IKC  & 29.7129 & 0.8709 & 28.4862 & 0.8449 & 29.4764 & 0.8254 & 28.2687  & 0.7914 \\
DAN & \underline{30.0293} & \underline{0.8813} & \underline{28.5829} & \underline{0.8589} & \underline{29.5708} & \underline{0.8269} & \underline{28.3362} & \underline{0.7993} \\
RealSR & 28.8903 & 0.8256 & 27.2931 & 0.8023 & 28.7308 & 0.8081 & 26.8275 & 0.7692 \\
Real-ESRGAN & 28.7841 & 0.8179 & 26.9721 & 0.7993 & 27.0132 & 0.7707 & 25.4315  & 0.7419 \\
SwinIR & 28.9764 & 0.8378 & 27.7532 & 0.8043 & 28.6215 & 0.8012 & 27.1922 & 0.7891\\
SR3 & 25.6895 & 0.7525 & 24.8917 & 0.7315 & 25.2515 & 0.7402 & 23.9709 & 0.7059\\
DFD-DCC & 28.7357 & 0.8104 & 27.8532 & 0.8073 & 26.6801 & 0.7821 & 25.4221 & 0.7137 \\
MDA-SR  & 28.7601 & 0.8647 & 27.0821 & 0.8290 & 27.7921 & 0.8082 & 26.8732  & 0.7671\\
RGT  & 28.9527 & 0.8755 & 27.8601 & 0.8575 & 29.5512 & 0.8253 & 28.3097 & 0.7899 \\
\textbf{Ours} & \textbf{32.0349} & \textbf{0.8933} & \textbf{30.8082} & \textbf{0.8748} & \textbf{30.8707} & \textbf{0.8370} & \textbf{29.6483} & \textbf{0.8113}\\
\bottomrule
\end{tabular}
}
\label{tab:results}
\end{table}

\begin{figure}[t]
\centering
  \begin{minipage}{0.247\textwidth}
        \small
        \centering
        \fboxsep=4pt  
        \fboxrule=0pt  
        \fbox{\textbf{HR}}
    \end{minipage}%
  \hfill
   \begin{minipage}{0.247\textwidth}
        \small
        \centering
        \fboxsep=4pt  
        \fboxrule=0pt  
        \fbox{\textbf{IKC}}
    \end{minipage}%
  \hfill
   \begin{minipage}{0.247\textwidth}
        \small
        \centering
        \fboxsep=4pt  
        \fboxrule=0pt  
        \fbox{\textbf{DAN}}
    \end{minipage}%
  \hfill
   \begin{minipage}{0.247\textwidth}
        \small
        \centering
        \fboxsep=4pt  
        \fboxrule=0pt  
        \fbox{\textbf{SRGAN}}
    \end{minipage}%
  \hfill

  \vspace{0.5pt}   
  \centering
  \begin{minipage}{0.247\textwidth}
    \scriptsize
    \includegraphics[width=\linewidth]{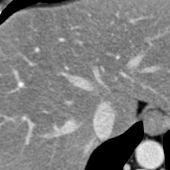}
  \end{minipage}%
  \hfill
  \begin{minipage}{0.247\textwidth}
   \scriptsize
    \includegraphics[width=\linewidth]{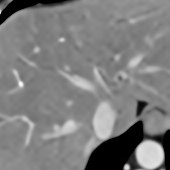}
  \end{minipage}%
  \hfill
  \begin{minipage}{0.247\textwidth}
    \scriptsize
    \includegraphics[width=\linewidth]{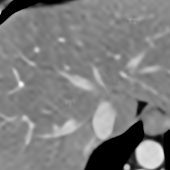}
  \end{minipage}%
  \hfill
  \begin{minipage}{0.247\textwidth}
    \scriptsize
    \includegraphics[width=\linewidth]{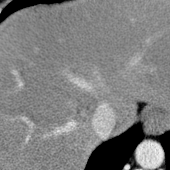}
  \end{minipage}%
  \hfill

  \vspace{0.5pt}   
  \centering
  \begin{minipage}{0.247\textwidth}
    \scriptsize
    \includegraphics[width=\linewidth]{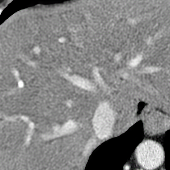}
  \end{minipage}%
  \hfill
  \begin{minipage}{0.247\textwidth}
   \scriptsize
    \includegraphics[width=\linewidth]{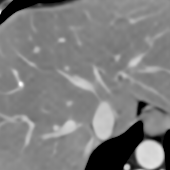}
  \end{minipage}%
  \hfill
  \begin{minipage}{0.247\textwidth}
    \scriptsize
    \includegraphics[width=\linewidth]{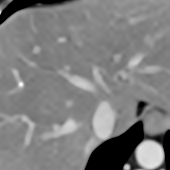}
  \end{minipage}%
  \hfill
  \begin{minipage}{0.247\textwidth}
    \scriptsize
    \includegraphics[width=\linewidth]{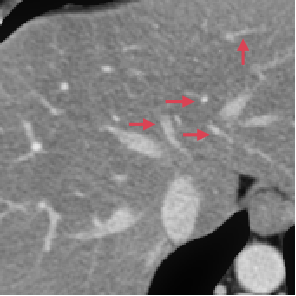}
  \end{minipage}%
  \hfill

  \vspace{0.5pt}
  \begin{minipage}{0.247\textwidth}
        \small
        \centering
        \fboxsep=4pt  
        \fboxrule=0pt  
        \fbox{\textbf{Real-ESRGAN}}
    \end{minipage}%
  \hfill
   \begin{minipage}{0.247\textwidth}
        \small
        \centering
        \fboxsep=4pt  
        \fboxrule=0pt  
        \fbox{\textbf{SwinIR}}
    \end{minipage}%
  \hfill
   \begin{minipage}{0.247\textwidth}
        \small
        \centering
        \fboxsep=4pt  
        \fboxrule=0pt  
        \fbox{\textbf{RGT}}
    \end{minipage}%
  \hfill
   \begin{minipage}{0.247\textwidth}
        \small
        \centering
        \fboxsep=4pt  
        \fboxrule=0pt  
        \fbox{\textbf{Ours}}
    \end{minipage}%
  \hfill
  
\caption{Qualitative results on the 3D-IRCADb dataset with a scale factor of 4. The top-left image shows a zoomed-in region of a high-resolution image, focusing on the liver and its internal structures for better visualization. Subsequent images display super-resolution results from various methods. The bottom-right image presents our proposed method's output, with red arrows highlighting regions where our approach generates more accurate and detailed structures.}
\label{fig:demo}
\end{figure}

\subsection{Comparison with State-of-the-Art Methods}
We compare our method against twelve state-of-the-art approaches: SRGAN~\cite{srgan}, EDSR~\cite{edsr}, ESRGAN~\cite{esrgan}, RCAN~\cite{rcan}, IKC~\cite{ikc}, DAN~\cite{dan}, RealSR~\cite{realsr}, Real-ESRGAN~\cite{real-esrgan}, SwinIR~\cite{swinir}, SR3~\cite{sr3}, DFD-DCC~\cite{dual-guidance}, MDA-SR~\cite{mda-sr}, and RGT~\cite{rgt}. Table~\ref{tab:results} demonstrates our framework's superior performance across both datasets and scaling factors. On the Pancreas dataset, our method surpasses the second-best approach by 2.0056$\text{dB}$/0.012 and 2.2253$\text{dB}$/0.0159 in PSNR and SSIM for $\times$2 and $\times$4 scaling factors, respectively. Similarly, for the 3D-IRCADb dataset, we achieve improvements of 1.2999$\text{dB}$/0.0101 and 1.3121$\text{dB}$/0.012 in PSNR and SSIM across both scales compared to the next best method.
\par
Fig.~\ref{fig:demo} presents a qualitative comparison between our method and competing approaches. Our results demonstrate superior artifact suppression, particularly in edge regions and internal structures, while other methods suffer from artifacts and blurred structures. Furthermore, our approach produces semantically accurate and detail-rich reconstructions, preserving clear edges, fine details, and textural structures.

\subsection{Ablation Studies}
We conduct comprehensive ablation studies to validate each component of our framework.

\subsubsection{Text Prompting Analysis}
We evaluate the impact of text-based guidance by comparing our full model against variants with omitted text prompts or modified LLaVA-Med instructions. Results in Table~\ref{tab:text} show that removing text prompts significantly degrades performance, highlighting their crucial role in providing valuable prior knowledge. Notably, our framework demonstrates robustness to minor variations in LLaVA-Med instructions, maintaining consistent performance---a crucial feature for practical applications.

\begin{table}[t]
\caption{Ablation study on the Pancreas dataset comparing different instructions to LLaVA-Med. \ding{55}: no text prompts; \ding{171}: using instruction \textit{Describe the anatomical structures in this CT image of the abdomen}; \ding{169}: using instruction \textit{List the major anatomic structures in this CT image of the abdomen}.}
\setlength{\tabcolsep}{4mm}
\centering
\small
\begin{tabular}{c|cccc}
\toprule
& \multicolumn{2}{c}{$\times$2} & \multicolumn{2}{c}{$\times$4}  \\ 
\multirow{-2}{*}{Instruction}  & PSNR & SSIM & PSNR & SSIM \\ \midrule
\ding{55} & 29.9012 & 0.8728 & 28.3015 & 0.8505 \\
\ding{171} & 32.0163 & 0.8928 & 30.7891  & 0.8724 \\
\ding{169} & \textbf{32.0349} & \textbf{0.8933} & \textbf{30.8082} & \textbf{0.8748} \\
\bottomrule
\end{tabular}
\label{tab:text}
\end{table}

\subsubsection{Visual Encoder}
We examine two configurations of the pre-trained VAE encoder from Stable Diffusion: learnable versus frozen. Results in Fig.~\ref{fig:ablation} indicate superior performance with the learnable configuration, likely due to its ability to adapt to medical image-specific features.

\subsubsection{Condition Network}
\begin{figure}[t]
\includegraphics[width=\textwidth]{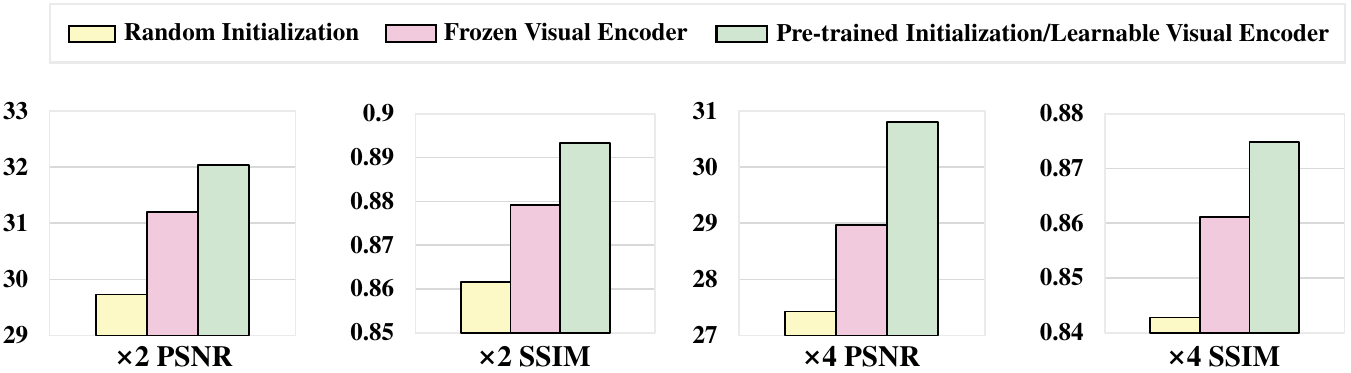}
\caption{Ablation study results on the Pancreas dataset. We evaluate the following configurations: random initialization (yellow) vs. pre-trained initialization (green), and frozen visual encoder (pink) vs. learnable visual encoder (green), for both $\times$2 and $\times$4 upscaling factors. Performance is measured using PSNR and SSIM. Our method consistently achieves the best results across all experimental settings.}
\label{fig:ablation}
\end{figure}

We investigate two initialization strategies for the condition network: pre-trained weights from Stable Diffusion's UNet versus random initialization. Fig.~\ref{fig:ablation} shows that pre-trained initialization yields better results in both PSNR and SSIM. This suggests that leveraging pre-trained weights enables more effective integration of conditional signals from low-resolution medical images and text prompts, ultimately leading to improved high-resolution image generation.

\section{Conclusion}
In this work, we present a novel framework for CT blind super-resolution that successfully adapts Stable Diffusion to the medical imaging domain. By using LLaVA-Med for semantic description generation and introducing a side-controlling strategy, we effectively bridge the domain gap between natural and medical images. Our experimental results validate the framework's effectiveness, showing consistent improvements over existing methods.
\par
The success of our approach suggests that pre-trained generative models can be effectively adapted for specialized medical imaging tasks. Looking ahead, the framework could be extended to other imaging modalities, and the integration of more sophisticated anatomical priors could further improve performance. We believe our work represents a step toward achieving high-quality CT imaging at reduced radiation doses, with potential impact on patient care through safer diagnostic procedures.
\par
Future work could explore extending this framework to other medical imaging modalities such as MRI and ultrasound to further validate its generalizability in various healthcare applications.


%
%
%
%

\end{document}